\documentstyle[12pt]{article}
\begin{document}

\begin{flushright}
IITM-TH-94-03\\
\vspace{0.2cm}
June 1994~~~~~~~~
\end{flushright}
\vspace{1cm}
\baselineskip=24pt
\begin{center}
{\Large \bf {Quantum Field Theory without divergences:\\
Quantum Spacetime}}\\
\vspace{0.5cm}
{G.H.Gadiyar}\\
{Department of Mathematics, Indian Institute of Technology,}\\
{Madras 600 036, INDIA.}
\end{center}
\vspace{1cm}
\noindent {\bf Abstract:} A fundamental length is introduced into
physics in a way which respects the principles of relativity and quantum
field theory. This improves the properties of quantum field theory:
divergences are removed. How to quantize gravity is also indicated.
When the fundamental length tends to zero the present version of quantum
field theory is recovered.

\vspace{0.5cm}
\noindent PACS 11. General Theory of fields and particles.
\newpage

In this paper the problem of divergences in quantum field theory is
solved. This is done by a simple modification of the concept of spacetime.
Mathematically what is being done is called a deformation [1].  The
idea is to introduce a fundamental length into physics. There are
compelling reasons to do this. Masters like Heisenberg, Dirac, Landau and
Schwinger [2]
have on various occasions talked about the logical inconsistencies
of quantum field theory. Further the problem of quantizing gravity
remains [3]. Wheeler [4] talks of the possibility of fluctuations at the
Planck length.

The origin of this work is as follows. Weinberg [5] has shown that
the equivalence principle and the General Theory of Relativity can be
derived from special relativity and quantum field theory. However the
problem of renormalization remains. Thus the author was tempted to play
with the axioms underlying the concepts of spacetime and pursue the
consequences. The article contains no exhaustive references to the
literature. This is because the results were arrived at without reference
to the literature and attempts to give references would be misleading.

Mathematicians call the process of introducing a fundamental constant
into a structure as deformation. Thus relativity which introduces the
velocity of light $c$ and quantum mechanics which introduces the Planck
constant $\hbar$ into the older structures, namely Galilean relativity
and classical mechanics, are called deformations.In each case the
two structures will be formally and mathematically similar to the older
structures though conceptually severe changes are needed.

Let us attempt to deform quantum field theory. It is immediate that the
Planck length $L$ is the only parameter available. To introduce the
parameter let us introduce a pair of operators $\hat{x}_\mu$ and
${\hat{x}_\mu}^{\#} $ satisfying the commutation relations
$$
[\hat{x}_\mu , {\hat{x}_\nu}^\#]~=~i\eta_{\mu \nu} L^2 \ .
$$
Further  the interval is now defined as
$$
\hat{s}^2~=~ \frac{1}{2} \left (\hat{x}_\mu \eta ^{\mu \nu} \hat{x}_\nu
{}~+~ {\hat{x}_\mu}^\# \eta ^{\mu \nu}{\hat{x}_\nu}^ \# \right ) \ .
$$

{}From the traditional algebra of operators it follows that the interval
is quantized in units of $L^2$. Further the commutation relations are
compatible with the other relation namely,
$$
[\hat{x}_\mu , \hat{p}_\nu]~=~i \hbar \eta _{\mu \nu} \ .
$$
But there is one problem: we have doubled the number of spacetime
coordinates and further we would like to have classical spacetime
in the limit of $L \rightarrow 0$. We therefore construct coherent
states of the annihilation operator constructed from $x^\mu$ and
$x^{\mu \#}$. To halve the number of variables we take those coherent
states $ |x>~=~ |x^\mu+ix^{\# \mu}>$ for which $x^\mu~=~x^{\# \mu}$.
It can now be checked that $<x|y>~=~e^{\displaystyle{-\frac{(x-y)^2}{2L^2}}}$
for
these states. This corresponds to taking a subset of the overcomplete
set.

To fix ideas take only one coordinate $x$. Then
$$
[\hat x, \hat{x} ^\#]~=~ i L^2
$$
and
$$
\hat{s}^2 ~=~ \frac{1}{2} \left (\hat {x}^2 ~+~ \hat {x}^{\#^2} \right).
$$
Set
\begin{eqnarray*}
d & = & \frac{\hat{x}~+~i\hat{x}^\#}{\sqrt{2} L}\\
d^{\dagger} & = & \frac{\hat{x}~-~i\hat{x}^\#}{\sqrt{2} L}\\
\left [d,d^{\dagger}\right ] & = & 1\\
\hat{s}^2 & = & \frac{L^2}{2}(dd^{\dagger}~+~d^{\dagger}d) \ .
\end{eqnarray*}
Now comes the halving of coordinates. We cannot let $\hat x$ equal $\hat
{x}^\#$
as operators. So we set $ x ~=~ x^\# $ on the coherent states which we
assume as the state of spacetime. Then
\begin{eqnarray*}
|x>& = & \displaystyle{e^{-\frac{x^2}{2L^2}}}\,\displaystyle{\sum ^\infty _0
\frac{[(\frac{1+i}{\sqrt{2}})\frac{x}{L} ]^n}{\sqrt{n!}} |n>}\\
<x|x'> & = & e^{\displaystyle{-\frac{(x-x')^2}{2L^2}}} \ .
\end{eqnarray*}
We normalize differently to
$$
<x|x'>~=~ \frac{1}{\sqrt{2 \pi}L} e^{\displaystyle{-\frac{(x-x')^2}{2L^2}}} \ .
$$
This goes over to $\delta (x-x')$ as $L \rightarrow 0 $ .The reason
for doing this will become clear shortly.

We now go over to quantum field theory. We have so far introduced two
ideas. We have doubled the number of coordinates to introduce a fundamental
length and changed the definition of interval.
This leads to
quantization of the interval. We have next introduced the idea that
spacetime is in a coherent state and halved the number of coordinates which
finally appear. These two ideas leave us with a fine balance between
discreteness and continuity as the following analysis will show.

Thus we see that functions of $x$ will now be functions of annihilation
operators acting on states $|x>$. This is just like in optics. To make
contact with quantum field theory is our next task.

Since there are spacetime fluctuations, a little thought shows that we
should take the action and average over the spacetime fluctuations.
Thus we are led to replace the action for the scalar field
$ \int \partial _\mu \phi \, \partial ^\mu \phi \, d^4x$ by
\linebreak $\int d^4x \, d^4y \, \partial _\mu \phi (x) \, K(x-y) \, \partial
^\mu \phi (y)$ where
$$K(x-y)~=~\left ( \frac {1}{\sqrt{2 \pi}L} \right )^4 e^{-
\displaystyle{\frac{(x-y)^2}{2L^2}}} \ .$$ This is got by sandwiching of the
operator corresponding to
$\partial _\mu \phi \partial ^\mu \phi $ between the states
$<x|$ and $|y>$ and averaging over the fluctuations.
It will turn out that the factor $K(x-y)$ will appear in several places.

To carry out the path integral will involve replacing
$$
Z~=~ \int {\cal D} \phi \ e^{-\displaystyle{\frac{1}{2} \partial \phi \partial
\phi ~+~ J \phi}}
$$
where $$
\partial \phi \,\partial \phi ~=~ \int d^4x \,\partial _\mu \phi (x) \,\partial
^\mu \phi (x)$$ and $$ J \phi ~=~ \int d^4x \,J(x) \,\phi (x) $$
by
$$
Z~=~ \int {\cal D} \phi \ e^{-\displaystyle{\frac{1}{2} \partial \phi K
\partial \phi ~+~ J K \phi}}
$$
where
$$
\partial \phi \,K \,\partial \phi ~=~ \int d^4x \,d^4y \,\partial _\mu \phi (x)
\,K(x-y) \,\partial ^\mu \phi (y)$$ and $$ JK \phi ~=~ \int d^4x \,d^4y\, J(x)
\,K(x-y)\,\phi (y) .$$ Thus the propagators in momentum space
$\displaystyle{\frac{1}{p^2}}$ are replaced by
$\displaystyle{\frac{e^{-\frac{p^2L^2}{2}}}{p^2}}$. This corresponds to a
regularization
of the theory. It is remarkable that this means something very simple in
the parametric representation [6] of the propagator.
\begin{eqnarray*}
\displaystyle{\frac{e^{- \frac{p^2L^2}{2}}}{p^2}}& =& \displaystyle{e^{-\frac
{p^2L^2}{2}}\int^\infty _0 e^{-\alpha p^2} d\alpha} \\
& = & \displaystyle{\int ^\infty _0 e^{-(\alpha~+~\frac{L^2}{2})p^2} d
\alpha}\\
& = & \displaystyle{\int ^\infty _{\frac{L^2}{2}} e^{-\alpha p^2 }d \alpha}
\end{eqnarray*}
by replacing $ \alpha + \frac{L^2}{2}$ by $\alpha$ and changing limits
of integration. Thus the theory is automatically regularized. To do
quantum electrodynamics is now simple.
It turns out that the propagator and vertices will be modified. However
the gauge principle survives. Nonlocal field theory thus arises naturally
from our assumptions. The parametric representation of Feynman amplitudes
can be utilized with a natural cutoff. Thus no new techniques
need be invented for calculations. Further the results of QED will be
reproduced but now with a natural cutoff.

We recapitulate what has happened. We assumed that spacetime is in a
coherent state and averaged over the fluctuations and this caused
the propagators to be regulated. Thus a natural regularization comes
about. Further the parametric representation makes the calculations
very simple. This is because we have deliberately deformed the original
theory and such results are bound to appear.

To quantize gravity is now fairly simple. The work of Weinberg [5] shows
the way. However with the structure of spacetime modified the infinities
which plague the older theory will go away. The spin-two graviton will
couple to a modified energy momentum tensor which will have nonlocal
structure. Physically one can say that point particles are replaced
by objects which are roughly of the order of $L$ in size due to the
new structure of spacetime.

Notice that modifying the action leads to modified classical equations of
motion as well. Thus the hope that Dirac expressed of modifying the
classical and quantum theory together is fulfilled in an unexpected way.

There are several conceptual issues involved. The reason is as follows.
Whenever one deforms an older theory the newer theory will have
mathematically similar structure. Witness the fact that the
Lie algebra structure survives in both classical and quantum mechanics.
Thus the language of the original theory and its deformation will be similar
but
conceptual issues will be thorny. The author refrains from any discussion
of conceptual issues as this is a luxury he cannot afford at present.

\vspace{0.5cm}

\noindent {\bf Acknowledgements:}
Thanks to Prof.E.C.G.Sudarshan for frequently emphasizing the structural
similarity between classical and quantum mechanics and the beauty of
coherent states.  Thanks to Dr.H.S.Sharatchandra for
short, illuminating discussions in the initial stages of the work.
Thanks to L.Kannan of PPST Foundation and Shambu Prasad of Dastkar for
encouraging new modes of thinking.

The author wishes to dedicate this work to Paramahamsa Yogananda whose
centenary is being celebrated.

\newpage

\noindent {\bf References}
\begin{enumerate}

\item {\it On the Relationship between Mathematics and Physics},
L.D.Faddeev, Asia Pacific Physics News vol.3, June-July,1988,
pp.21-22.

\item  {\it Niels Bohr and the Development of Physics}, W.Pauli
Ed.\\
{\it Theoretical Physics in the Twentieth century: A
memorial volume to Wolfgang Pauli}, M.Fierz and V.Weisskopf Eds.\\
{\it Inward Bound }, A.Pais.\\
{\it Selected Papers on Quantum Electrodynamics},
J.Schwinger Ed.
\item {\it Quantum Theory of Gravity. Essays in honour of the
sixtieth birthday of Bryce S.DeWitt}, S.M.Christensen Ed.
\item {\it Gravitation}, C.W.Misner, K.S.Thorne and J.A.Wheeler\\.
{\it Magic without Magic. John Archibald Wheeler},
J.R.Klauder Ed.
\item {\it Photons and Gravitons in S-Matrix Theory: Derivation
of Charge Conservation and Equality of Gravitational and Inertial
Mass},S.Weinberg, Phys.Rev B1049-1056 vol.135,4B, 1964.\\
{\it Photons and Gravitons in Perturbation Theory:
Derivation of Maxwell's and Einstein's Equations}, S.Weinberg,
Phys.Rev B988-1002, vol.138, 4B, 1965.
\item {\it Quantum Field Theory}, C.Itzykson and J.B.Zuber.
\end{enumerate}

\end{document}